# DIGITAL GLOBAL PUBLIC GOODS

Johan Ivar Sæbø, University of Oslo, johansa@ifi.uio.no

Brian Nicholson, University of Manchester, brian.nicholson@manchester.ac.uk

Petter Nielsen, University of Oslo, pnielsen@ifi.uio.no

Sundeep Sahay, University of Oslo, sundeeps@ifi.uio.no

**Abstract:** The purpose of this paper is to define and conceptualize digital global public goods (DGPGs) and illustrate the importance of contextual relevance in ICT4D projects. Recent studies have examined the importance of digital artefacts with public goods traits, emphasizing the significant potential for socio-economic development. However, we know little about the theoretical and practical dimensions of how we can align the public goods traits of such artefacts to create relevance in the context they are implemented. To address this gap we review the literature firstly to develop a definition and conceptual basis of DGPGs and then to illustrate the importance of relevance: how to align DGPGs with context to meet local needs. The illustration draws from a case study of the District Health Information systems (DHIS2). The paper advances both the theoretical and practical understanding of DPGs in development processes.

## 1. INTRODUCTION

National socio-economic development processes are becoming increasingly intertwined with digital technologies. Digital technologies are flexible and multipurpose by being reprogrammable and based on combinable and reusable components. They are standard based to handle a variety of digital contents and to allow innovation by the many. Many low and lower middle-income countries (LLMICs) have positioned the digital at the center of their reform agendas (Heeks, 2020). In the increasingly complex and interconnected development landscape, digital public goods (DPGs) are being positioned as the core platform on which to build the digital solutions upon. The UN Secretary-General's High-level Roadmap for Digital Cooperation defines digital public goods as "open-source software, open data, open artificial intelligence models, open standards and open content that adhere to privacy and other applicable international and domestic laws, standards and best practices and do no harm" (United Nations, 2020, p. 35). The same report states that "digital public goods are essential in unlocking the full potential of digital technologies and data to attain the Sustainable Development Goals, in particular for low- and middle-income countries" (United Nations, 2020, p. 8). A range of associated organizations and initiatives adopt similar definitions, sometimes with a specific normative addition that DPGs should help attain the Sustainable Development Goals (SDGs). One such organization is the Digital Public Goods Alliance, a multi-stakeholder initiative to provide a one-stop-shop on information about DPGs. Digital Square, another such multi-stakeholder initiative, seeks to be a "marketplace" of DPGs for health as "investment opportunities" for global health organizations. These initiatives are influential in setting priorities for digital technologies in international development and raising concerns around ethics, development agendas and funding concerns.

A limitation of the above definitions relates to the simplistic tendency to equate free and open source software with DPG status, which we argue over-emphasizes the production side of the DPG, including the processes of design, development and distribution, whilst not paying adequate attention to the demand side and how the DPG meets local needs. To address this gap in the literature, this paper sets out with two aims. The first concerns building an improved understanding and definition of the nature of a digital public goods, which we do by proposing the term Digital Global Public Goods (DGPGs), drawing on concepts of public goods and global public good and unpacking an understanding of the digital in development. Second, we illustrate this definition by drawing on the case of the District Health Information System (DHIS2) software over the last 2 decades.

The remainder of the paper is organized as follows: in the next section we develop a definition of DGPG based on discussing relevant literature concerning DPGs. In the following section, we provide a case study and analysis of DHIS2, a digital platform for public health in developing countries (see www.dhis2.org). We frame our analysis of DHIS2 as a DGPG and base our discussion on published works related to DHIS2 and on the authors' ongoing and long-term theoretical and practical engagement with the design, development, and implementation of DHIS2

## 2. DIGITAL GLOBAL PUBLIC GOODS

In this section, we build a definition and conceptual framework of DGPGs, drawing from three related stream of research - public goods, global goods and digital technologies.

Public goods are non-rivalrous and non-exclusive. Non-rivalry means that consumption by one individual does not subtract what is available for others to consume, and non-excludability means that one cannot prevent anyone from using the good (or it is prohibitively expensive to do so). The concept of public goods originates from economics (Samuelson, 1954) where a prevailing focus is on the undersupply of public goods in a free market. Without prospects of payment, there will be limited economic incentives to provide the good, despite the high aggregate benefit to society. Some coordination of collective action to supply them is necessary, which is typically the realm of the public sector. Typical examples of public goods in the literature include infrastructure (lighthouses and streetlights), environment (clean air and disease control), and public services (immunization, fire stations and national defense). The production of public goods results in positive externalities, i.e. benefits for third parties that did not agree to consume and pay for the good. For example, lighthouses will benefit all seafarers, their families, and customers of the transported goods too. Immunization will lead to herd immunity, which is beneficial also for those not vaccinated. While public goods are by default non-exclusive, technological change may alter this by bringing down the cost of enforcing payment. An example of this is introduction of cost-efficient automatic toll roads where previously the road was free to use.

The concept of global public goods (GPG) represent an approach to capture the global challenges in enabling access to public goods that transcend national borders (Kaul, 2013). It puts emphasis on the related costs and benefits and the accessibility to public goods across geographies, social and economic groups, gender, and generations (Kaul et al., 1999). This concept is widely applied across many areas related to socio-economic development including the environment, international financial stability, peace and security, human rights (Long & Woolley, 2009) as well as the area of global health (Moon et al., 2017; Smith & MacKellar, 2007).

GPGs take an important role in discussions on global agendas, goals and the establishment of governance structures to create incentives for different groups to contribute with their share (Kaul et al., 1999). Even if public goods are global and not produced by a single nation, they can only be realized by several countries taking policy initiatives on national levels and through international cooperation (Kaul et al., 2003). However, critics have raised a concern that there is a tendency to uncritically apply the concept of global public goods to the level that it loses its meaning and becomes a 'buzzword' to attract funding (Smith & MacKellar, 2007), which makes the term incoherent and abstract (Long & Woolley, 2009).

How may *digital* global public goods be conceptualized? Digital technologies (such as open source software or digital platforms) have qualitatively different characteristics than other technologies. By virtue of their nature, they are relatively easier to circulate across time and space compared to technologies with physical properties e.g. infrastructure and machines. This may make digital technologies easy and cheap to replicate and share, even globally (Yoo et al., 2010). "Digital" therefore implies the capability to re-program, modularize and recombine, build upon, and share digital goods, potentially enabling their appropriation and modification to build relevance in multiple local contexts. This flexibility allows digital technologies to have generativity (Zittrain, 2008) reflected in the success of commercial digital platform ecosystems such as those around iOS and Android, reaping the benefits of positive network effects (Tiwana, 2013). However, Sahay (2019) points out that appropriation of the digital in multiple contexts is not an unproblematic given, but shaped by various "distortions" such as related to knowledge, capacity and local politics. Understanding the nature of these distortions and how to engage with them, helps shape local relevance.

We have so far briefly introduced public goods, global public goods and digital technologies. Public goods points towards market failure and the need for collective action, global public goods does the same on a global scale, and digital technologies have the nature of being inherently flexible for appropriation in a local context. DGPGs represents a combination of these defining criteria. First, they are designed to be available to anyone free of charge without any license costs. Second, they are relevant in local contexts on a global scale by not being prohibitively difficult to learn how to use, implement, and appropriate. We thus adopt a wide understanding of accessibility, to include digital goods being understood, capable of being influenced, capable of being adapted and appropriated. The notion of accessibility thus relates not just to ease of acquisition, but more importantly how easy they are for users to understand, utilize and make relevant in context. There are mutual dependencies between these constituent parts, shaped broadly by the process and capabilities of making the goods locally relevant. In a self-reinforcing cycle, global use adds the contextual diversity necessary to develop for global relevance, but also raises the challenge of the goods becoming too generic and offering a 'design from nowhere' (Suchman, 2002). An open source application that is prohibitively difficult to re-program, adjust, or localize, becomes less of a DGPG. A DGPG can as such be understood as a whole which is more than the sum of its parts. We summarize our perspective of DGPGs in the table below:

| *Digital Global Public Goods are digital goods designed as non-rivalrous, non-excludable, locally relevant on a global scale.* | |
| --- | --- |
| **Public Goods** | Non-rivalrous and non-excludable by design |
| **Digital** | Re-programmable, modularized and re-combinable |
| **Global** | Relevant locally on a global scale |
| **DGPG as more than the sum of the parts** | Relevance on a global scale can only be achieved with adaptability, provided by non-excludable, non-rivalrous, digitally modifiable technologies. Due to these traits, the technologies also display positive network effects, increasing the benefits with increasing scale. |

*Table 1 A definition of Digital Global Public Goods (DGPG)*

## 3. CASE STUDY: DHIS2 AS A DGPG

The above definition and how the traits may combine into DGPGs are demonstrated in this section illustrated with prior literature on the DHIS2 health information systems platform. We selected a case study as they are "particularly useful in the early stages of research on a new topic, when not a lot is known about it" (Myers, 2001, p. 89). While the history of DHIS2 is reported in many research papers (Adu-Gyamfi et al., 2019; Braa et al., 2007; Roland et al., 2017) and more than 50 PhD theses, we draw upon Braa and Sahay's (2012) description of the transitions in history of the initiative, which we also updated to current time. To highlight local relevance, we identified and focus on 3 key sets of processes: i) participation of users; ii) building of capacity and local support mechanisms;

and, iii) improving flexibility of the DHIS2 platform. Arguably, these aspects contribute to the development of local relevance, which we define as "the capabilities of the local context to effective use the DGPG platform to meet local needs, while contributing to the global scalability of the platform."

We first provide a brief overview of the background of the DHIS2 initiative, ongoing over 2 decades, and then discuss the aspects of local relevance over the four identified phases of its evolution.

### 3.1. A brief overview of the DHIS2 initiative

The DHIS2 initiative anchored within the broader Health Information Systems Programme (HISP) research and development initiative of the University of Oslo, Norway (UiO), has its origins in the South African post-apartheid health reform efforts. The first version of the software (DHIS1) was developed in South Africa and was transformed into the web based DHIS2 platform in 2006. It has slowly grown in maturity with new capabilities and improved functionalities; attracted global attention and funding, and experienced increasing adoption in more than 70 countries in the Global South (see www.dhis2.org). At the time of writing, DHIS2 has the status of *de facto* global standard for health information system development and is supported by a consortium of global partners (e.g. WHO, Norad, Global Fund, UNICEF amongst others). The design, development and dissemination of DHIS2 is coordinated by the Department of Informatics at University of Oslo (UiO) under the Health Information Systems Programme (HISP), where it represents a key focus of research and education.

### 3.2. Phase 1 (1995-2001): Local development for local use

Processes of participation: A key component of the African National Congress (ANC) government's post-apartheid reform agenda was the strengthening of the country's HIS, particularly in making it more decentralized and integrated. The process of reform was driven by a team of South African activists returning to the country from exile and researchers from UiO immersed in the Scandinavian approach to participatory design and action research. Emancipation of the field level health workers through decentralization was the underlying agenda guiding the technology development process (Braa & Hedberg, 2002). This was the birth of the HISP initiative with the priority to establish an integrated and decentralized health system, and the development of district health information system software became the focal point and the Version 1 (DHIS1) was born. The development progressed as a highly participatory and multi-disciplinary process, driven primarily by national nurses and doctors, who would give requirements and the development team would respond with rapid prototypes, often on a daily basis. Braa and Sahay (2012) describe this process as the technology was used as Lego bricks where the DHIS1 development and reconstruction of the health system took place as mutually reinforcing and inextricably intertwined processes.

Process of capacity building and local support mechanisms: An NGO HISP South Africa was born which was anchored within the School of Public Health at the University of Western Cape. Capacity building was at the core of the local support provision, as the university initiated various short term summer and winter courses, many of which were run by HISP South Africa team members, and thousands of health staff attended the courses (Braa et al., 2004). Additionally, the HISP South Africa team would carry out in-service training for health staff in their work contexts. The software development took place from a base in Western Cape, with strong linkages established between the development, implementation and user teams. The capacity building and support mechanisms can be summarized as taking place in the mode of "for and in the local context" of the country.

Technology development process: The first version of DHIS was open source and built using Visual Basic and Microsoft Access and implemented based on standalone installations in the health facilities. The development team was located at the University of Western Cape (UWC) and consisted of two core software developers and a group of HISP members acting as mediators between users and the developers. Based on rapid prototyping cycles in close collaboration with users in selected health facilities, the software matured, it was scaled to provincial level and finally as a national standard in all districts. The software developers and public health implementers and users worked closely together in a small and agile team, and their ongoing interactions contributed to the rapid scaling of the systems, till it became a national standard by 2001.

In summary, with respect to local relevance, the initiative was strong in processes of participation, capacity building and local support. While the software was locally relevant in South Africa, it was not globally scalable, as was evidenced in the next phase of evolution.

### 3.3. Phase 2 (2001 to 2008): Exploring potential of scaling of DHIS1

Processes of participation: Two sets of processes were at the core of shaping participation in this phase. One, there was a process of taking the DHIS1 from South Africa to multiple other countries such as Mozambique, Tanzania, Ethiopia, Malawi, India and Cuba. Participation processes were attempted in different (from South Africa) political contexts and resulted in systemic challenges such as the top down and hierarchical political structure of Cuba that did not accept the bottom up and activist approach of South Africa, leading to the HISP initiative to be abruptly terminated (Sæbø & Titlestad, 2004). Funded by Norad, UiO worked on establishing strong research and education programmes in collaboration with national universities and Ministries of Health (MoH). For example, 6 faculty members from the Universidade Eduardo Mondlane (UEM) in Mozambique enrolled for the PhD programme at UiO working in an action research mode on topics of relevance for the MoH. Simultaneously, a Masters programme in Health Informatics was established in collaboration with UiO. Similar models of engagement were established in Tanzania, Ethiopia and Malawi. These changes firstly strengthened three way links between UiO, the local universities and the MoHs, and the action research mode provided a more academic mode to participatory processes.

Process of capacity building and local support mechanisms: The onus of organizing and managing these processes largely shifted to the settings of the universities, with students at the core within the framework of their research projects. This shift came with its particular strengths and weaknesses. The obvious strength was this process sought to strengthen national institutional processes of tertiary education and health information capacity in the MoH. Such strengthening would contribute to enhanced sustainability of efforts and systems. A weakness of this model was that unlike South Africa where an NGO was responsible for supporting DHIS implementation, and could rapidly respond to user needs, in Mozambique for example, the support was bound in various layers of bureaucracy of their respective universities and the MoH. As a result, support could not be provided in time in relation to the urgency required. This greatly adversely affected the implementation outcomes.

Technology development process: The point of departure for this process was the DHIS1 developed in South Africa. The process of translating this to the country contexts met with multiple design-reality gaps (Nhampossa, 2005). For example, there was the issue of language, as for example to Portuguese in Mozambique. There were different underlying logics of working, for example while in South Africa with the agenda of empowerment and "local use of information", a primary focus was on the development and use of pivot tools, in Mozambique the users wanted ready-made report generation functionality rather than having to develop it themselves. The technology support for these countries still came from South Africa, and because of the limitations of a standalone system, laptops would need to be sent from Mozambique to South Africa, who would then install the new version incorporating required fixes and send it back to Mozambique. As can be imagined, this was a very cumbersome and time-consuming process, and did not contribute effectively to local technical capacity development.

In summary, with respect to local relevance, longer term and institutionalized processes of participation and capacity building were arguably established, but unlike in South Africa, it was inadequate to meet the short-term implementation needs. The DHIS1 was found to be inherently non-scalable. The result of this was in Mozambique, Ethiopia and Cuba, the HISP projects were terminated by 2005-06, while in other contexts like India and Tanzania limited pilot projects were initiated.

### 3.4. Phase 3: (2008 to 2012): Transition to the web-based version

<u>Processes of participation</u>: The initial years of this transition involved building participation around the development of the web-based version DHIS2. Two key conditions shaped this participation process. One, the development process shifted from South Africa to the Department of Informatics at UiO and was now driven primarily by Masters and PhD students. They were then both geographically and culturally distant from the contexts of use. Two, the strong multi-disciplinary approach born in South Africa, was replaced by a primarily technical approach with focus on novel rather than appropriate technologies, which implied losing the public health anchoring. For example, the DHIS2 was based on a stack of Java based technologies, skills for which was relatively limited in many developing countries. There were also infrastructure constraints. For example in the Ethiopian national university, there were severe internet constraints, which meant the local developers could not download the new builds released by the UiO developers. Processes of mutual participation of the UiO and country teams was severely constrained during this process of new technology development.

<u>Process of capacity building and local support mechanisms</u>: Many of the doctoral candidates from countries started to complete their PhDs and go back to their countries and take up positions in their universities or MoH. This helped to strengthen institutional capacities while also participating in teaching of the Masters programmes now established and running well. Some new countries like Sri Lanka and Malawi also established their respective programmes, slowly leading to the development of more global health informatics capacities made relevant to their country contexts. Some other countries, like India, Vietnam and Bangladesh established local NGOs (called HISP groups) to support local implementations, and developed their own means and mechanisms of participation guided by various factors such as political contexts, available resources and capacities, and the agreements they had with their respective governments. While a diversity of participatory processes grew during this period, a common theme was it was becoming increasingly technical and distant from users. The UiO team's engagement in participation was now largely mediated by the country support structures.

<u>Technology development process</u>: The DHIS2 was born in the state of Kerala in India, jointly developed and piloted in one clinic by a team of UiO and HISP India developers. The web-based nature of DHIS2 made it immediately attractive to governments, and 3-4 states in India and other countries (Sierra Leone and Kenya) initiated processes to pilot and implement DHIS2 in their contexts. As a team of developers from Oslo relocated to Kenya to carry out development in context, some major breakthroughs were achieved, for example Kenya became the first African country to achieve a national roll out of a web based HIS (Manya et al., 2012). There was a rapid evolution of the functionalities of DHIS2 making it more amenable to both local customization and also their global scaling. For example, Gizaw et al. (2016) advanced the concept of "open generification" using examples of how locally developed functionalities (such as for mortality reporting in Ethiopia and category combinations in Tajikistan) could become part of the DHIS2 core and made available to the world at large. The growth of DHIS2 caught the attention of global partners, such as WHO and Global fund, and processes of supporting DHIS2 development were initiated leading to a rapid increase in the uptake of DHIS2 in countries.

In summary, local relevance was some ways compromised with the change in the character of participation, but the use of more modern technologies enhanced its use in multiple local contexts. Local support mechanisms through HISP groups was strengthened as increasingly governments started to rely on the DHIS2, placing pressure on its continued maintenance and support.

*Sæbø et al.*                                                                                                          *Digital Global Public Goods*## 3.5. Phase 4: (2013 to current): Dealing with high global and national scaling

<u>Processes of participation</u>: Participation was now organized primarily at two levels. One, country teams (HISP groups) organized their own processes of participation in respective country contexts, with mixed results. While South Africa continued their relative mature modes of participation through secondment of a number of their staff to provincial and national governments, other countries showed relatively reduced levels of engagement with users and a stronger focus on building more technical skills on the DHIS2, which was evolving at a high speed. Two, the UiO team largely engaged with countries through the mediating HISP groups rather than directly with users, which was now no longer possible given the scale of operations (Roland et al., 2017). Further, the models of funding were also rapidly changing, as till around 2012 nearly all development was carried out by students supported through research council funds, now different donor money was being obtained. An implication of this shift was that the priorities of the donors tended to be emphasized over country voices, reducing the impetus on engaging user participation.

<u>Process of capacity building and local support mechanisms</u>: About ten HISP groups were established around the developing world (for example, in India, Mozambique, Uganda, Tanzania, West Africa, Vietnam) and there was increasing focus (and funds) to strengthen their capacities, primarily around the customization and use of DHIS2. A key mechanism for this was through the UiO supported "DHIS2 Academies" arranged periodically in different regions on topics of system development, implementation, server management, app development and so on (since 2011, around 100 academies have had around 5000 participants). The DHIS2 Academies are arranged by the HISP groups in their regions, sometimes supported by UiO core team members. In 2017, a free online academy is also offered on "Fundamentals of DHIS2", and more such courses are in the process of development. There is also an Annual DHIS2 Conference, where representatives from NGOs, MoHs and researchers share learnings and showcase innovation and best practices. These have contributed to building a vibrant global community of practice and network of practitioners around DHIS2. These activities are at the same time at a distance from users in developing countries. This challenge has been recognized by UiO and various measures are being undertaken to reduce this gap. For example, recently "regional HISP groups" were established in South Asia, East Africa and West Africa, to enable pooling of regional capacities so as to better organize country support.

<u>Technology development process</u>: The technology development processes have rapidly moved towards increased professionalization and specialization, from the early days when students conducted the development. Today there is a professional team of more than 30 developers, distributed globally, and organized on lines of a modern software house with different teams including front-end and back-end. To provide support for the implementation and use in a variety of context and for different use-cases, DHIS2 is based on the newest of technologies and a platform architecture. DHIS2 is a Java-based web application and runs on multiple platforms, and is interoperable with other relevant applications in the domain. It is available in many languages and has support for local contribution to a repository of translations. Participation is also promoted by an online community platform, mailing lists, source code repositories, and issue tracker.

With its generic core and platform architecture, DHIS2 provide interfaces that allows the development of apps to suite local needs and integration with other systems. From being a tool primarily used for collection, aggregation and presentation of aggregate health data, the platform has facilitated new use cases including patient management and individual records. This versatility is also illustrated by the implementation and use of DHIS2 for health commodity logistics management and in agriculture and education. However, taking advantage of the openness and flexibility of the platform requires competence and human capacity (Msiska & Nielsen, 2018). There are different initiatives to support participation including the development of different kinds of boundary resources, such as component libraries, tutorials, and documentation.

In summary, in this phase the focus has been on building global rather than local relevance, which is inevitable given the scale and associated funding mechanisms. The ongoing challenge is to

Proceedings of the 1st Virtual Conference on Implications of Information and Digital Technologies for Development, 2021

896

manage this global scale, while ensuring the successful principles of engagement and support learnt in the South African experience are not lost, but revitalized in the fast changing context.

We have in this case study presented three phases of the development of DHIS2, and below we summarize this in table 2 along three dimensions of local relevance: Participation, capacity building and local support, and technology development

| Phase | Local Participation | Capacity building and local support | Technology development |
|---|---|---|---|
| 1<br>1995-2001 | Situated, multi-disciplinary | Local NGO, local university anchoring | Situated, contextualized, custom-made |
| 2<br>2001-2008 | Trending towards universities as other modes of participation faced systemic challenges | University-based, vulnerable to university priorities and timescale | Desktop version did not scale, increased demand for web-based solutions |
| 3<br>2008-2012 | Increasing distance between designers and users | Emergence of local and regional support groups ("HISP groups"), establishment of stronger university programs, sometimes by exchange PhD students | Shift to web-based DHIS2. Generification to address increasing scale |
| 4<br>2013-current | Centered on HISP groups. Scale and increased global funding challenge participation further | DHIS2 Academies, growing of a community supported by online fora, platforms, conferences. | Platformization. Professionalized core team and emergent distributed "third party" developers |

*Table 2 Characteristics of local relevance*

## 4. DISCUSSION AND CONCLUSION

In this paper we sought to build an improved understanding of digital public goods by proposing and defining the term DGPG, and examine the DHIS2 software and associated activities in light of this. The story of DHIS2 shows that the accomplishment of a DGPG is not an absolute, or a process that is ever finished or perfected, but an ongoing mission that takes place alongside ongoing striving for relevance to context. We focused the analysis of the case on three aspects of seeking relevance; local participation, capacity building and local support, and technology development, summed up in Table 2. Our main contribution is thus to shed light on the "demand" side of DGPGs, arguing that digital public goods cannot be made without taking the context into consideration.

The current open source license and the availability of DHIS2 points towards the accomplishment of DGPG status by being non-rivalrous and non-excludable. Its digital nature, of being re-programmable, modularized and re-combinable, is apparent from implementation by a variety of organizations (Ministries of health, NGOs, PEPFAR, MSF etc.) and for different use cases (ranging from routine aggregate data to patient tracking, agriculture and education). Implemented in more than 80 countries, it also shows relevance locally on a global scale. Thus, DHIS2 corresponds to the definition of a DGPG. The careful balance between generification (Gizaw et al., 2016) and local flexibility (Roland et al., 2017) has reaped positive network effects.

The accomplishment of DGPG status was complex and many challenges are still lingering. First, funding of the development is reliant on international donor agencies, and their priorities may change. Compared to more traditional, and material, public goods, the public good nature is not a default but may be changed for instance by the introduction of a license. Second, the tensions emerging with scaling and striving to be relevant both globally and locally, while serving the needs of an increasingly diverse user-base, challenges relevance (Nicholson et al., n.d. forthcoming, 2019). DHIS2 shows how the digital and the global nature of DGPGs are intertwined. For example, a platform architecture can allow both shared stable resources and flexibility for specific, local development (Roland et al., 2017). Careful work with boundary resources can lower the bar for local adaptations (Li & Nielsen, 2019). At the same time, the focus on the dual mission of making the technology a globally relevant platform to be used across different use-cases and domains, has implications on the relationship between the developers and the users. While the core developers started out with a very intimate relationship with the users and the context in the early 1990's, this is the case no more. At the same time, with the platform architecture of DHIS2, local expertise has the flexibility to implement solutions relevant in the local context. The local experts do also have the possibility to change DHIS2 by feeding new requirements, use-cases and innovations back to the core-team. However, this requires them to invest time in understanding whether this is a common requirement across context and describing the use-cases in a way acceptable by the core team. Such initiatives will also be considered in light of what else is on the roadmap, and finally decided by the core team.

The case of DHIS2 and its development over time shows the relationship between the constituents of DGPG and in particular the relationship between accessibility and local relevance. While accessibility in terms of users being able to download a software platform is relatively easy to achieve, developing the skillsets of the users to understand the software, implement it effectively and putting it to use in a way that leads to better decisions is a different story. We argue that the relative success of DHIS2 hinges not only on the qualities of the software and its abilities, but also the global efforts put into building local capacities and regional support mechanisms for its implementation and use.

The paper contributes theoretically by providing a novel definition and conceptualization of DGPG illustrated with the case of DHIS2. The definition contributes analytical support towards improved understanding of the role of DGPGs in socio-economic development in particular a recognition of the importance of the "demand side". A limitation of the paper is that we do not in depth discuss practical contributions for producers of DGPGs. While this will be the focus of further enquiry, we hope to direct attention to a renewed emphasis on a broader perspective on accessibility and relevance for users. Users and policy makers may draw on the lessons of the case illustrations when considering investment and implementation of DGPGs.